\documentclass[12pt,preprint]{emulateapj}
\usepackage{natbib}
\shorttitle{Time Dilation Signatures and Gamma-ray Bursts}
\shortauthors{D. Kocevski \& V. Patrosian}

\begin{document}

\title{On The Lack of Time Dilation Signatures in Gamma-ray Burst Light Curves}

\author{Daniel Kocevski \altaffilmark{1}}
\author{Vahe Petrosian \altaffilmark{2}}

\altaffiltext{1}{Kavli Institute for Particle Astrophysics and Cosmology, Stanford University, 2575 Sand Hill Road M/S 29, Menlo Park, Ca 94025 }
\altaffiltext{1}{Center for Space Science and Astrophysics, Varian 302c, Stanford University, Stanford, CA 94305}


\begin{abstract}

We examine the effects of time dilation on the temporal profiles of gamma-ray burst (GRB) pulses.  By using prescriptions for the shape and evolution of prompt gamma-ray spectra, we can generate a simulated population of single pulsed GRBs at a variety of redshifts and observe how their light curves would appear to a gamma-ray detector here on Earth. We find that the observer frame duration of individual pulses does not increase as a function of redshift as one would expect from the cosmological expansion of a Friedman-Lema\^{\i}tre-Robertson-Walker Universe.  In fact, the duration of individual pulses is seen to decrease as their signal-to-noise decreases with increasing redshift, as only the brightest portion of a high redshift GRB's light curve is accessible to the detector.  The results of our simulation are consistent with the fact that a systematic broadening of GRB durations as a function of redshift has not materialized in either the \emph{Swift} or \emph{Fermi} detected GRBs with known redshift.  We show that this fundamental duration bias implies that the measured durations and associated $E_{\rm iso}$ estimates for GRBs detected near an instrument's detection threshold should be considered lower limits to their true values.  We conclude by predicting that the average peak-to-peak time for a large number of multi-pulsed GRBs as a function of redshift may eventually provide the evidence for time dilation that has so far eluded detection.

\end{abstract}

\keywords{gamma rays: bursts --- galaxies: star formation}

\section{Introduction}

Gamma-ray bursts are among the most distance objects observed in the Universe.  Having been detected as nearby as z = 0.0085 for GRB~980425 \citep{Galama99} and as distant as z = 9.4 for GRB~090429B \citep{Cucchiara11} , the redshift range over which these objects occur spans almost the entire length of the observable Universe. This large dynamic redshift range, combined with the transient nature of their prompt gamma-ray emission, makes them excellent source with which to look for evidence of time dilation due to the expansion of the Universe.  At such large distances, the duration of prompt gamma-ray burst temporal profiles should be stretched by a factor of $1+z$ due to cosmological expansion in Friedman-Lema\^{\i}tre-Robertson-Walker (FLRW) models of the Universe, resulting in an increase in the average burst duration as a function of redshift.

\begin{figure}
\includegraphics[height=.23\textheight,keepaspectratio=true]{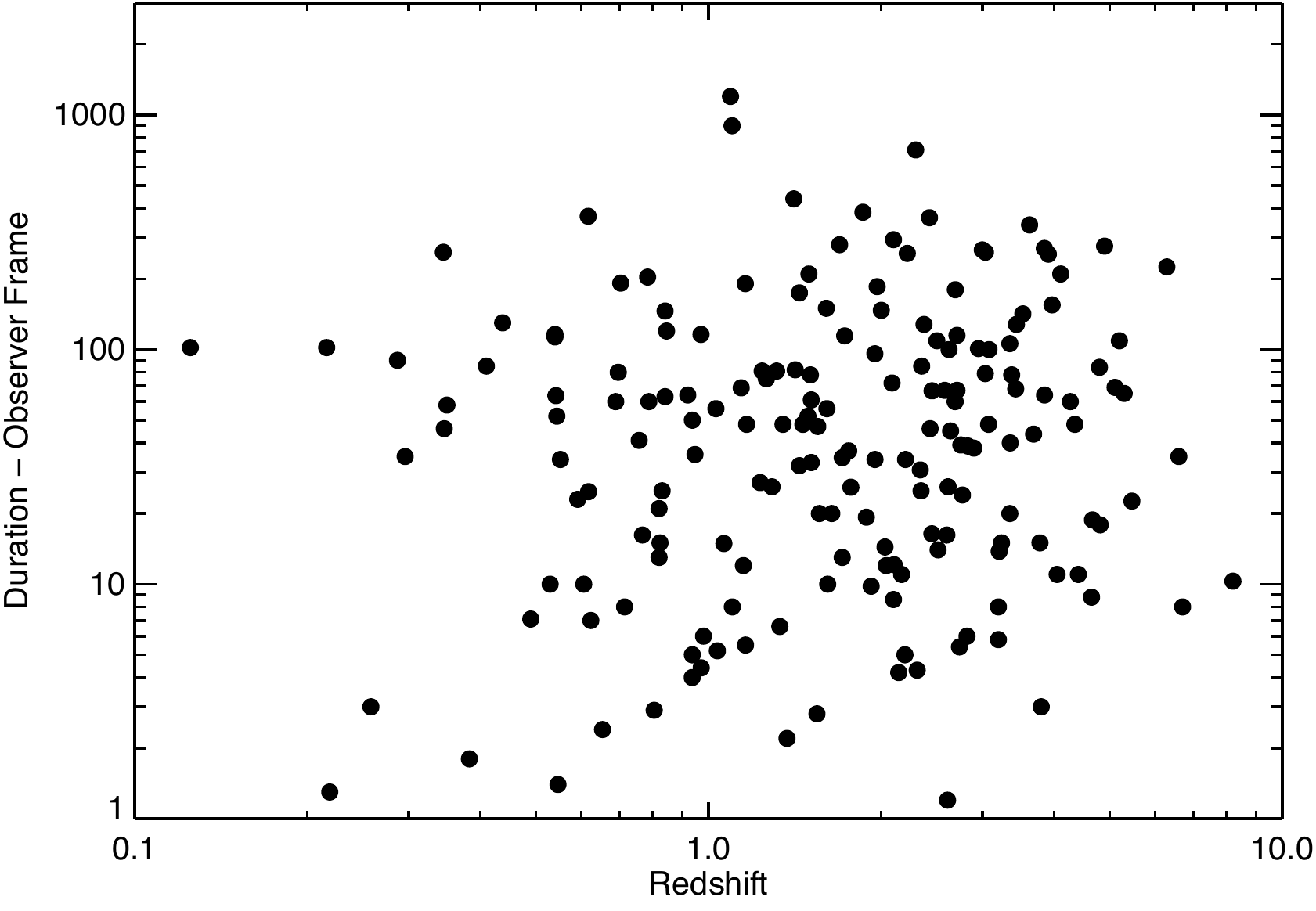}
\caption{T90 duration vs. redshift for 195  \emph{Swift} detected GRBs as reported via GCN.}
\label{Swift-T90VsRedshift}
\end{figure}

Such evidence for a systematic broadening of GRB durations as a function of distance has not materialized in either the \emph{Swift} \citep{Gehrels04} or \emph{Fermi} detected GRBs with known redshift \citep{Campana07}.  In fact, the most distance GRBs to date, GRB~980425 at z = 9.4, GRB~090423 at $z \sim 8.1$ and GRB~080913A at $z \sim 6.7$, have measured durations of 5.5 sec \citep{Cucchiara11}, 10.3 sec \citep{Salvaterra09} and 8.1 sec \citep{Greiner09} respectively, giving them rest frame durations of $\sim0.52$ sec, $\sim1.13$ sec, and $\sim1.04$ sec.  These values are far shorter than the average rest frame duration of $\sim 23-30$ sec for long GRBs detected at $z \sim 1$.  The lack of any correlation between a GRB's duration and its distance can be seen quite clearly for  \emph{Swift} detected GRBs with known redshift shown in Figure \ref{Swift-T90VsRedshift}.

In this paper, we examine the effects of time dilation on the temporal profiles of GRBs through the use of a new spectral evolution code that can model the observed properties of GRBs as a function of redshift.  By using prescriptions for the shape and evolution of prompt gamma-ray spectra, we can generate a simulated population of single pulsed GRBs at a variety of redshifts and observe how their resulting light curves would appear to a gamma-ray detector here on Earth.  We find that the observer frame duration of individual pulses does not increase as a function of $1+z$ as one would expect from cosmological time dilation.  In fact, the duration of individual pulses is seen to decrease as their signal-to-noise decreases with increasing redshift, as only the brightest portion of a high redshift GRB's light curve is accessible to the detector.  This effect is analogous to the problem faced when measuring the physical radii of galaxies at large distances, where the regions of lower surface brightness fall below the sensitivity of the collecting instrument.  In addition, the energy at which most of the photons are emitted (i.e$.$ the peak of the $\nu F_{\rm \nu}$ spectra) for high redshift bursts is redshifted to energies where traditional gamma-ray detectors become less sensitive, decreasing the observed signal even further.  The net result is that all estimates of duration and energetics for high redshift GRBs can only be taken as lower limits to their true values.  We conclude that GRB temporal profiles that exhibit narrow pulses separated by significant quiescent periods, may in fact the tell tail sign of high redshift GRBs and predict that the average peak to peak duration for a large number of multi-pulsed GRBs as a function of redshift may eventually provide the evidence for time dilation that has so far eluded detection.

We present an overview of our population synthesis code in Section 2.1, followed by a more in-depth description in Sections 2.2 through 2.4.  We present the result of our simulation in Section 3 and discuss the implications of our results in Section 4.

\section{GRB Model} \label{sec:Models}

Gamma-ray burst continuum spectra can evolve quite dramatically over the course of a burst.  This evolution is generally characterized by an overall softening of the spectra, with the peak of the ${\nu}F_{\nu}$ spectrum ($E_{\rm pk}$) evolving through the detector bandpass over the duration of the burst.  This evolution will be delayed by the effects of time dilation for GRBs at high redshifts, resulting in a longer observed spectral lag between the high and low energy channels and a broadening of the pulse profile.  At the same time, the observed GRB flux falls as a function of increasing luminosity distance.  The net effect is that soft and faint emission will become increasingly difficult to observe with traditional gamma-ray detectors, which suffer large drops in sensitivity at low energies.  The observed bursts properties are therefore a complex convolution of the effects of cosmological redshift and detector sensitivity and hence we turn to simulations to obtain a better idea of how these bursts would appear in the observer frame.  

\begin{figure}
\includegraphics[height=.23\textheight,keepaspectratio=true]{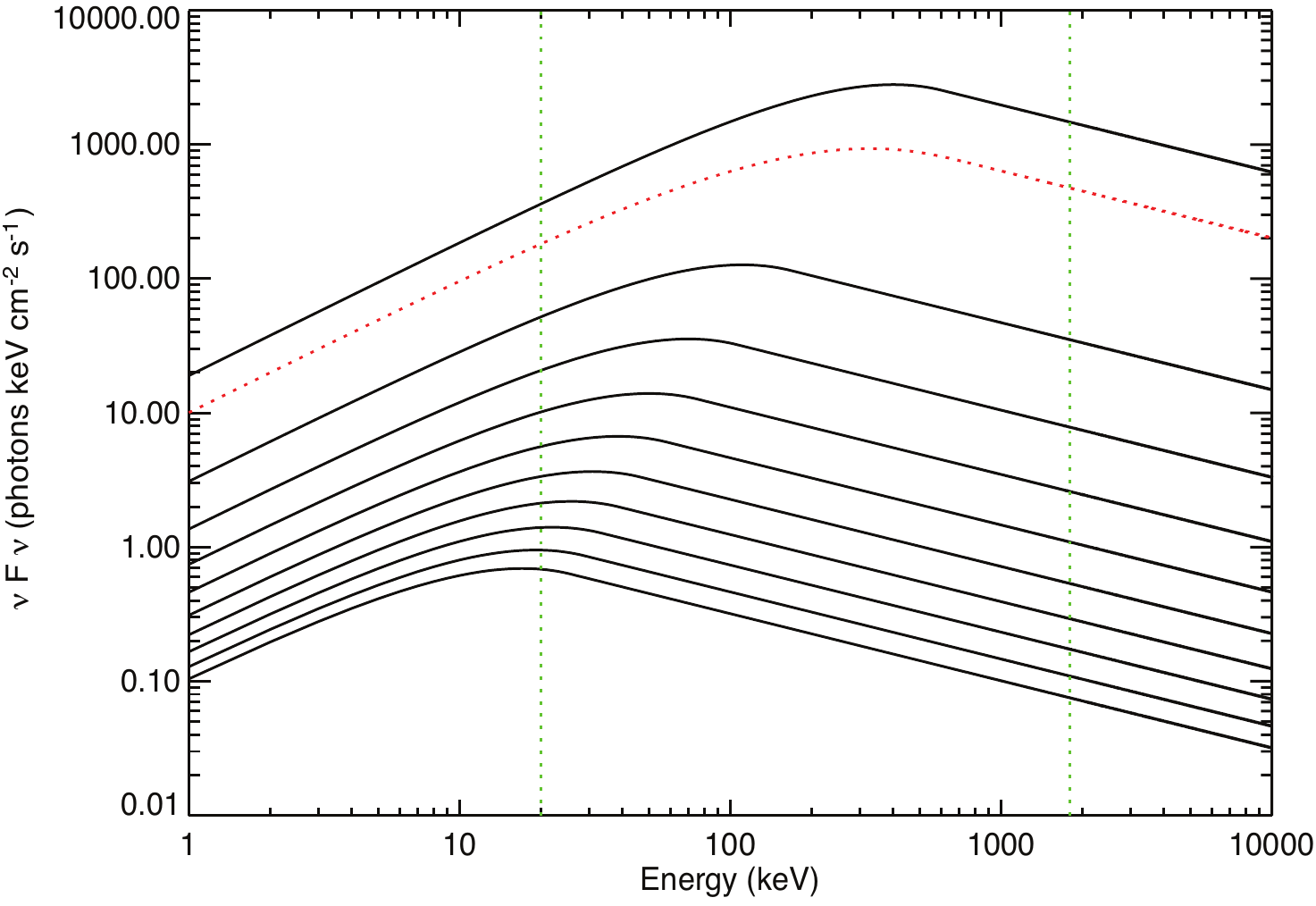}
\caption{The hard to soft spectral evolution of $E_{\rm pk}$ as a function of time as prescribed by relativistic kinematics for a spherical shell traveling towards the observer with high Lorentz factor.}
\label{SpectralEvolution}
\end{figure}

Two empirical correlations form the basis for the GRB model that we have constructed to investigate this question.  The first is the {\it hardness-intensity correlation} or HIC, which relates the instantaneous hardness of the spectra and the instantaneous energy flux $F_{E}$, within individual pulses \citep{Golenetskii83}. For the decay phase of a pulse, the most common behavior of the HIC is a power-law relationship between $F_{E}$ and the peak of the ${\nu}F_{\nu}$ spectrum, $E_{\rm pk}$, of the form  $F_{E}\propto E_{\rm pk}^{\eta}$ \noindent, where $\eta$ is the HIC power-law index. The second correlation is the {\it hardness-fluence correlation} or HFC \citet{Liang96} which describes the observation that the instantaneous hardness, or $E_{\rm pk}$, of the spectra decays exponentially as a function of the time-integrated flux, or fluence, of the burst.  The HFC can be stated as $E_{\rm pk} = E_{0} e^{-\Phi / \Phi_{0}}$, where $\Phi(t)$ is the photon fluence integrated from the start of the burst and $\Phi_{0}$ is the exponential decay constant.  

\begin{figure} 
\includegraphics[height=.23\textheight,keepaspectratio=true]{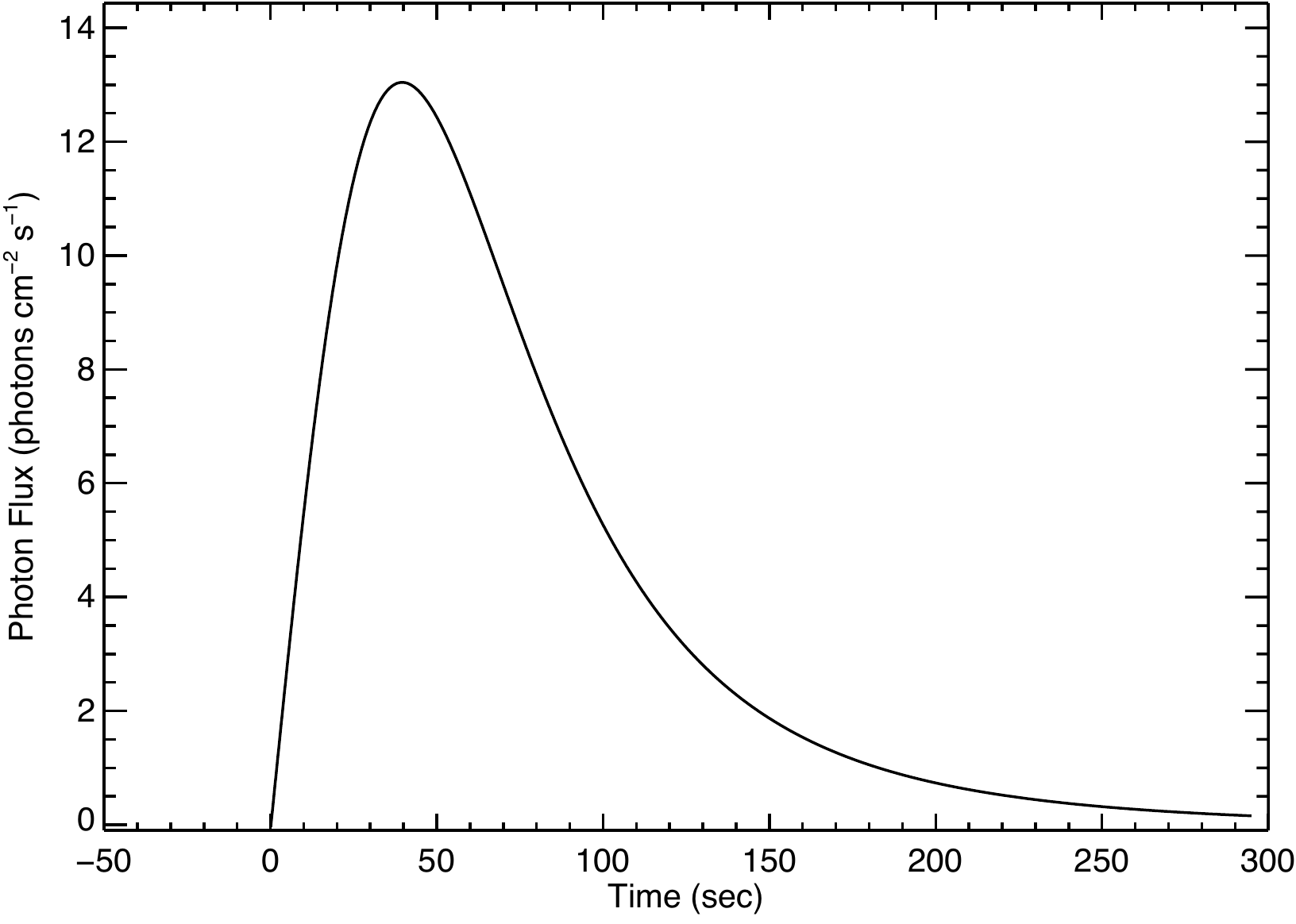}
\caption{The resulting fast rise exponential decay (FRED) pulse shape generated by the model by integrating the bolometric GRB spectrum at each time bin.}
\label{PhotonFluxLightCurve}
\end{figure}

\citet{Kocevski03} have shown that both correlations can be produced through simple relativistic kinematics when applied to a spherical shell expanding at relativistic velocity (i.e. the curvature of a relativistic shock front).  By assuming a spectral shape, in this case a Band model \citep{Band93}, for the GRB's instantaneous photon spectrum and evolving that spectrum in time according to these two empirical correlations, we are able to reproduce the Fast Rise Exponential Decay (FRED) pulse shape that is so ubiquitous in GRB data.  

An example of a time resolved photon spectra and the resulting photon flux light curve for a simulated burst using a Band model with an initial $E_{\rm pk,src} = 500$ keV and a fixed $\alpha = 1$ and $\beta = -2.2$ can be seen in Figures \ref{SpectralEvolution} and \ref{PhotonFluxLightCurve} respectively.  The photon flux light curve in Figure \ref{PhotonFluxLightCurve} is calculated by integrating the photon spectra over the instrument's energy range and as such is not bolometric.  Therefore the shape of the resulting light curve will be greatly effected by the location of $E_{\rm pk}$ within the instrument's energy range.

In order to convert our modeled photon spectrum into a count spectrum and eventually a count light curve, we take our simulated photon spectrum and fold it through an instrument data response matrix (DRM).  For the purposes of this analysis, we use a response file from the \emph{Burst and Transient Source Experience} (BATSE) \citep{Meegan92} that was generated for a real burst which occurred nearly at zenith for one of BATSE's Large Area Detectors (LAD).   The DRM describes the distribution of counts over the instrument's energy channels due to the arrival of a photon of a given energy.  The result is a time resolved count spectra as a function of channel energy and time.

\begin{figure*}
\begin{center}
        \includegraphics[scale=0.50]{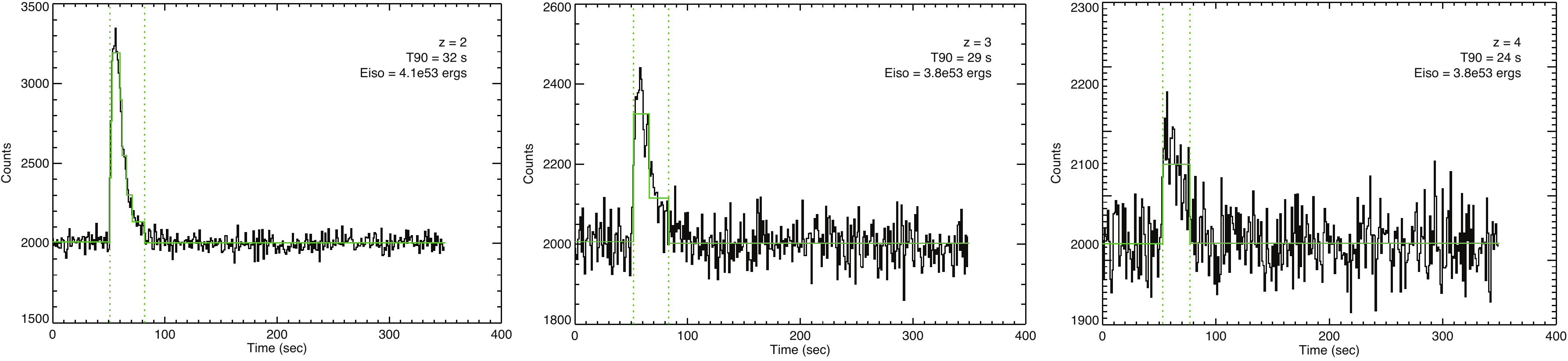}
\caption{The effects of increasing redshift on the observed light curve of a GRB pulse.  The effects of time dilation are hidden by the decreasing signal-to-noise and the pulse duration in fact decreases with increasing redshift.}
\label{CountLightCurves}
\end{center}   
\end{figure*}

We then add a Poisson distributed, energy dependent, background spectrum derived from the median backgrounds of a sample of BATSE detected bursts to each time resolved count spectra.  A count light curve with a realistic background spectrum can then be produced as a function of time by integrating the individual time resolved count spectra over the instrument's effective energy range for each time bin.  The burst duration is then measured using a Bayesian block algorithm to determine periods of emission that are above the Poisson background.  We refer to this estimate as the T100 duration (as opposed to the traditional T90 duration found through flux integration methods). 

We refer the reader to Kocevski et al.\ (2011) for a more comprehensive description of the GRB model used for the present analysis.

\section{Results} \label{sec:Results}

Figure \ref{CountLightCurves} shows the resulting count light curves for the simulated GRB shown in Figure \ref{SpectralEvolution} placed at three different redshifts, along with the Bayesian block reconstruction of the burst duration (solid green lines).  Contrary to what is expected due to cosmological time dilation, the duration of the simulated pulse decreases as a function of increasing redshift.  This is primarily due to the decreasing signal-to-noise of the observed pulse with increasing redshift, which progressively limits only the brightest portion of the GRB's light curve from being accessible to the detector.  This effect is analogous to the problem faced when measuring the physical radii of galaxies at large distances, where the regions of lower surface brightness fall below the sensitivity of the collecting instrument.  In addition, the energy at which most of the photons are emitted (i.e$.$ the peak of the $\nu F_{\rm \nu}$ spectra) is redshifted to energies at which the instrument becomes less sensitive, reducing the observed signal even further. 

\begin{figure} 
\includegraphics[height=.23\textheight,keepaspectratio=true]{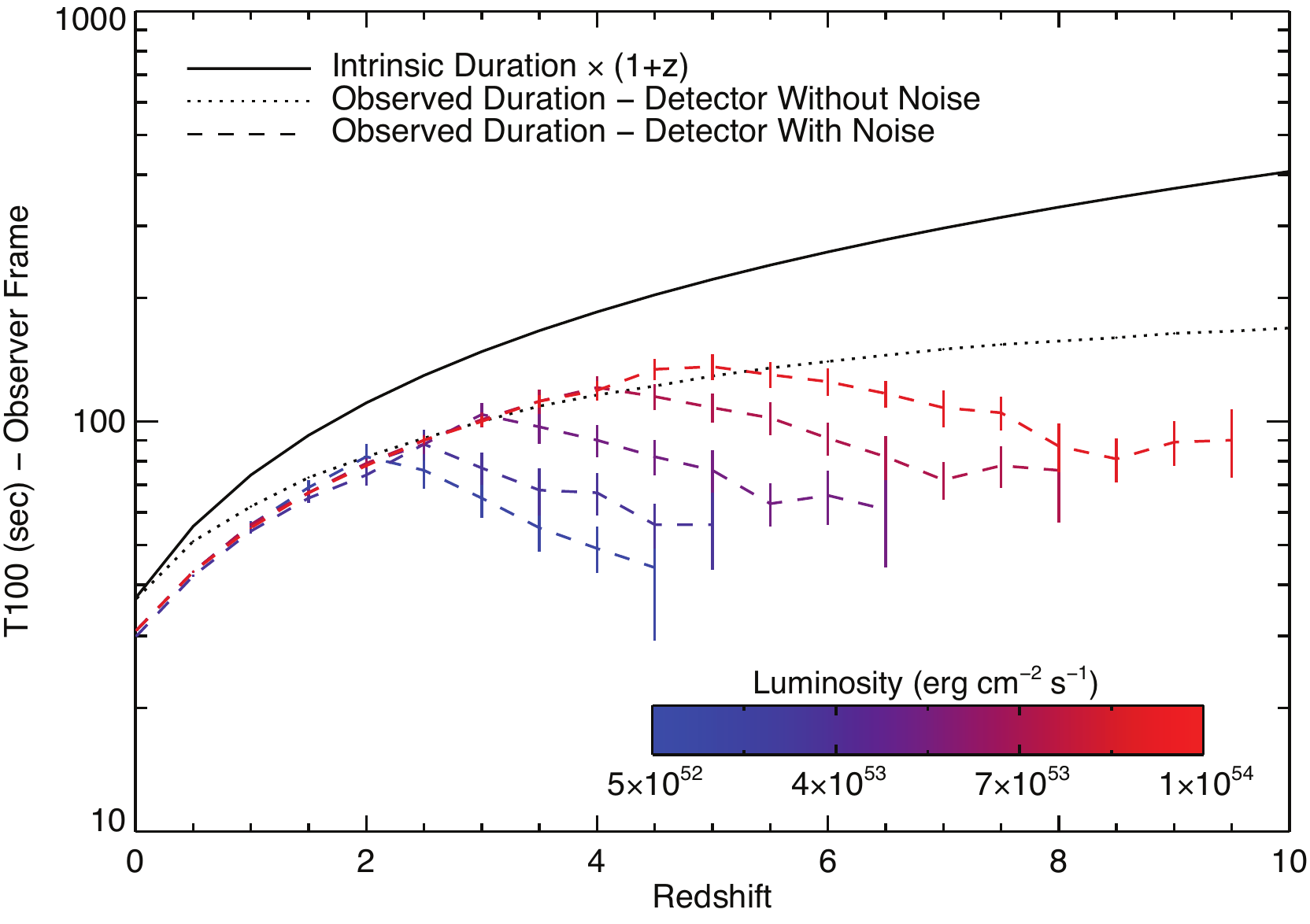}
\caption{The observer frame duration for a set of simulated GRB pulses, of equal intrinsic duration but varying intrinsic luminosity, as a function of redshift.  Even a perfect detector collecting data over a limited energy range would not faithfully measure the expected time dilation effects on a GRB pulse as a function of redshift.  When noise is added to the observed signal, the measured duration turns over and begins to decrease with increasing redshift.   }
\label{T100VsRedshift-ObserverFrame}
\end{figure}

Both of these effects can be seen in Figure \ref{T100VsRedshift-ObserverFrame}, where we plot the observer frame duration for a set of simulated GRB pulses, of equal intrinsic duration but varying intrinsic luminosity, as a function of redshift.  The solid line represents the expected observer frame duration in a FLRW Universe, found by multiplying the pulse's intrinsic duration by $(1+z)$.  The dotted line represents the observed duration as measured from the photon flux light curves, found by integrating the photon spectra over the detector's energy range.  The dashed lines represent the observed duration as measured from count light curves for each of the simulated GRB pulses.  These count durations actually represent the median value of 200 count light curve realizations for each redshift bin, computed in order to account for statistical fluctuations in the simulated background which may dominate the duration measurements as the observed signal-to-noise decreases with increasing redshift.  The error bars represent the standard deviation of the resulting duration distribution for each set of 200 realizations.

The divergence between the predicted observer frame duration and the duration of the photon flux light curve can be understood as due to the redshifting of the underlying GRB spectra.  Because the simulated photon flux light curves are non-bolometric, their shape will be greatly effected by the redshifting of $E_{\rm pk}$ towards the lower edge of the detector's energy range.   Essentially, the pulse profile that would have been seen by the detector if the burst had occurred at $z = 0.01$, for example, is not the same as the profile seen at $z = 2$, because the lower end of the GRB spectrum is no longer detected by the instrument.  Therefore, even a perfect detector that observes over limited energy range would not faithfully measure the expected time dilation effects on a GRB pulse as a function of redshift.

The divergence between the predicted observer frame duration and the measured duration becomes wider when considering duration measurements made through the use of the detector's count data. When noise is added to the observed signal, the measured duration turns over and begins to decrease with increasing redshift.  As the pulse's signal-to-noise falls, only the brightest portion of pulse becomes accessible to the detector, until the observed duration approaches zero and the pulse is no longer detected.  As seen in Figure \ref{T100VsRedshift-ObserverFrame}, the redshift at which the transition between a rising and falling duration depends largely on the luminosity of the pulse, but is also influenced by the burst's intrinsic $E_{\rm pk,src}$, since the redshifting of $E_{\rm pk,src}$ towards the lower edge of the detector's energy range acts to further reduce the signal-to-noise of the pulse.  

\begin{figure} 
\includegraphics[height=.23\textheight,keepaspectratio=true]{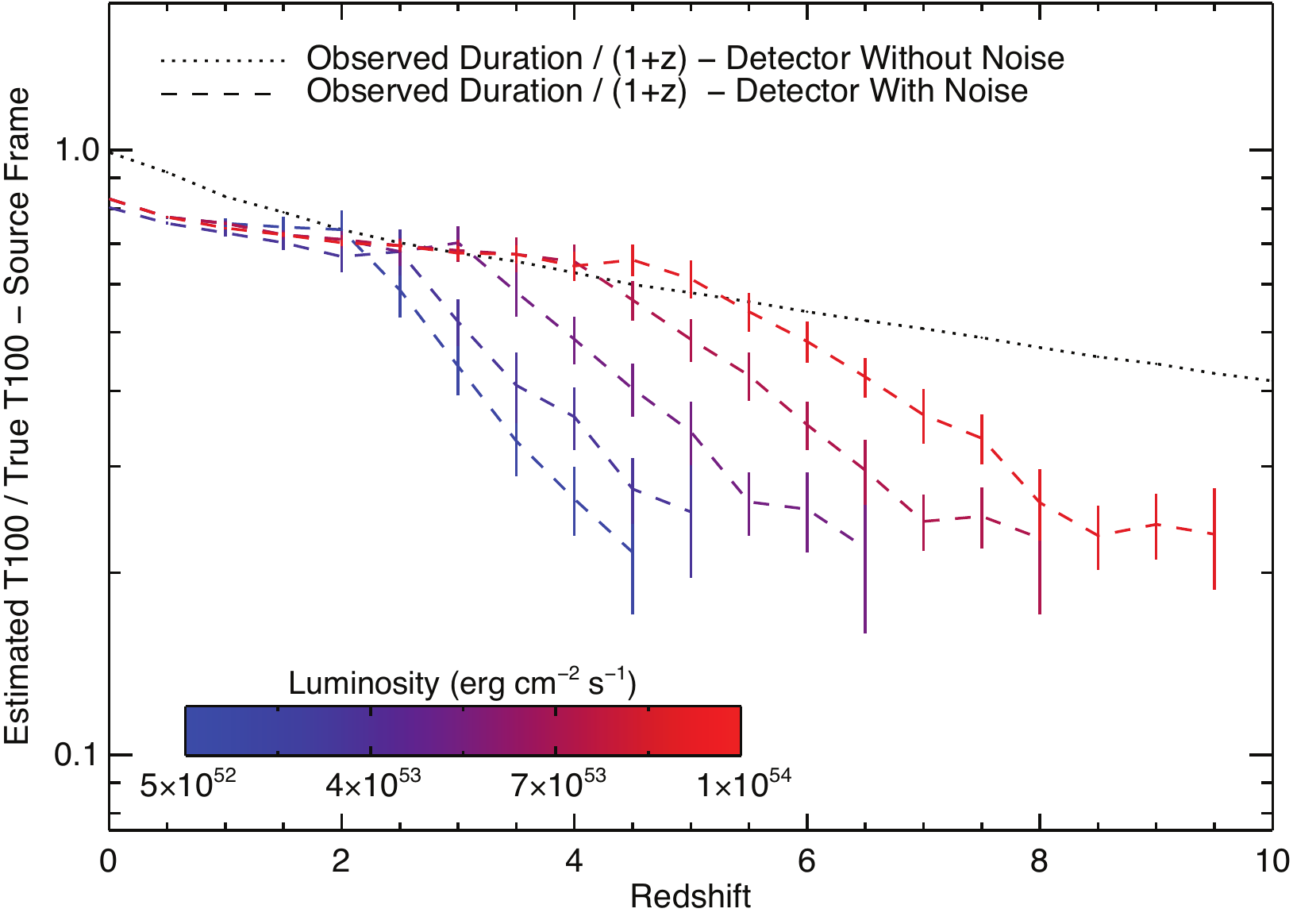}
\caption{The estimated source frame duration, normalized to the true intrinsic duration, for a set of simulated burst's of varying luminosity.  Transforming the under-estimated observer frame durations into the source frame has the effect of further widening the difference between the pulse's true intrinsic duration and our duration estimates}
\label{T100VsRedshift-SourceFrame}
\end{figure}

\begin{figure} 
\includegraphics[height=.23\textheight,keepaspectratio=true]{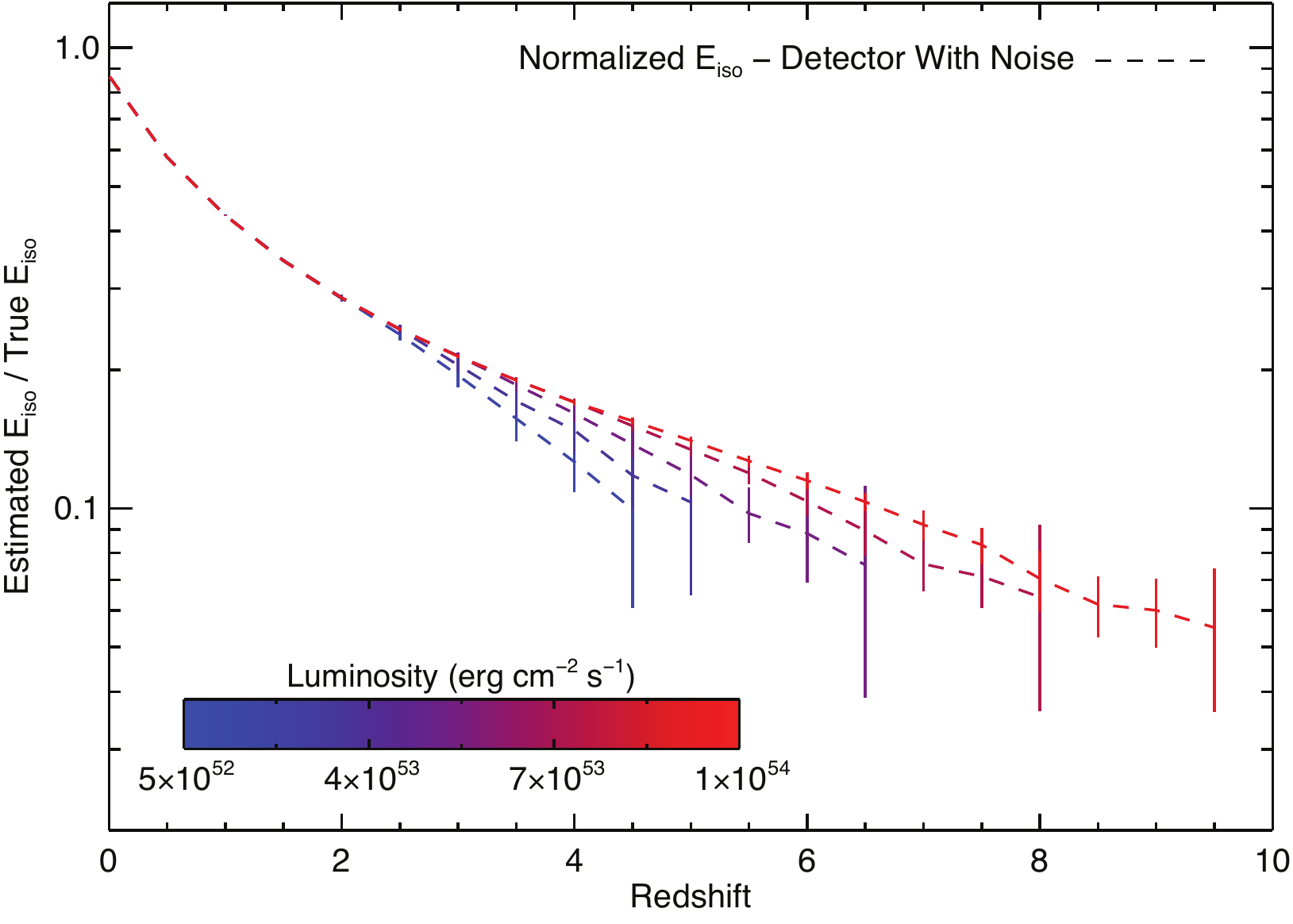}
\caption{The burst's estimate $E_{\rm iso}$ normalized to the burst's true $E_{\rm iso}$ for a set of simulated burst's of varying luminosity. The systematic error that the duration bias introduces  propagates into our estimate of the burst's energetics}
\label{EisoVsRedshift}
\end{figure}

Transforming the observed durations into the source frame by dividing by $(1+z)$ has the effect of further widening the difference between the pulse's true intrinsic duration and our estimates.  The systematic error that this duration bias introduces also propagates into our estimate of the burst's total energy release, as $E_{\rm iso}$ is estimated by integrating the burst's flux over the observed duration.  Figures \ref{T100VsRedshift-SourceFrame} and \ref{EisoVsRedshift} show both of these effects.  

In Figure \ref{T100VsRedshift-SourceFrame}, we have plotted the estimated source frame duration, normalized to the true intrinsic duration, for burst's of varying luminosity, but of equal intrinsic duration (dashed lines).  The turn over observed in Figure \ref{T100VsRedshift-ObserverFrame} can now be seen as a steeping of the estimated source frame duration as a function of redshift.  In all cases, the under-estimation of the source frame duration can approach as much as 80$\%$ near the detection threshold of the instrument.  

The consequences that this systematic error has on the estimates of $E_{\rm iso}$ can be seen in Figure \ref{EisoVsRedshift}, where the ratio of the the pulse's estimate to true $E_{\rm iso}$ can reach as much as 90$\%$, or more, as the pulse nears the detector's sensitivity threshold.  The $E_{\rm iso}$ measurements presented here were k-corrected to a standard energy range of $10-10000$ KeV, hence this error in our estimate of $E_{\rm iso}$ is largely due to the under-estimate of the duration over which the burst's observed flux is integrated.  

In Figure \ref{T100VsSNVsFlux}, we quantify the severity of the duration bias as a function of the observed signal-to-noise ratio by simulating a 4000 single peak GRBs at $z = 1$ of equal intrinsic duration, but varying luminosity.  The estimated observer frame duration normalized to the intrinsic duration is plotted vs the pulse's observed signal-to-noise, with a color coding displaying the peak photon flux of the observed signal.  Again, the bias in the duration estimate is largest for intrinsically weak bursts with low signal-to-noise as seen in the detector.  Although the bias is far less severe for the highest signal-to-noise bursts, the offset between the estimated duration and the true duration is still evident and due to the redshifting of $E_{\rm pk,src}$.  The divergence between the true observer frame duration and the estimated duration at low signal-to-noise ratios reflect the similar breaks seen in Figures \ref{T100VsRedshift-SourceFrame} and \ref{T100VsRedshift-SourceFrame}.  More importantly, this divergence occurs at roughly the same signal-to-noise ratio, in this case SNR $\sim 25$, for all of the bursts in this simulation.

\section{Discussion} \label{sec:Discussion}

\begin{figure} 
\includegraphics[height=.23\textheight,keepaspectratio=true]{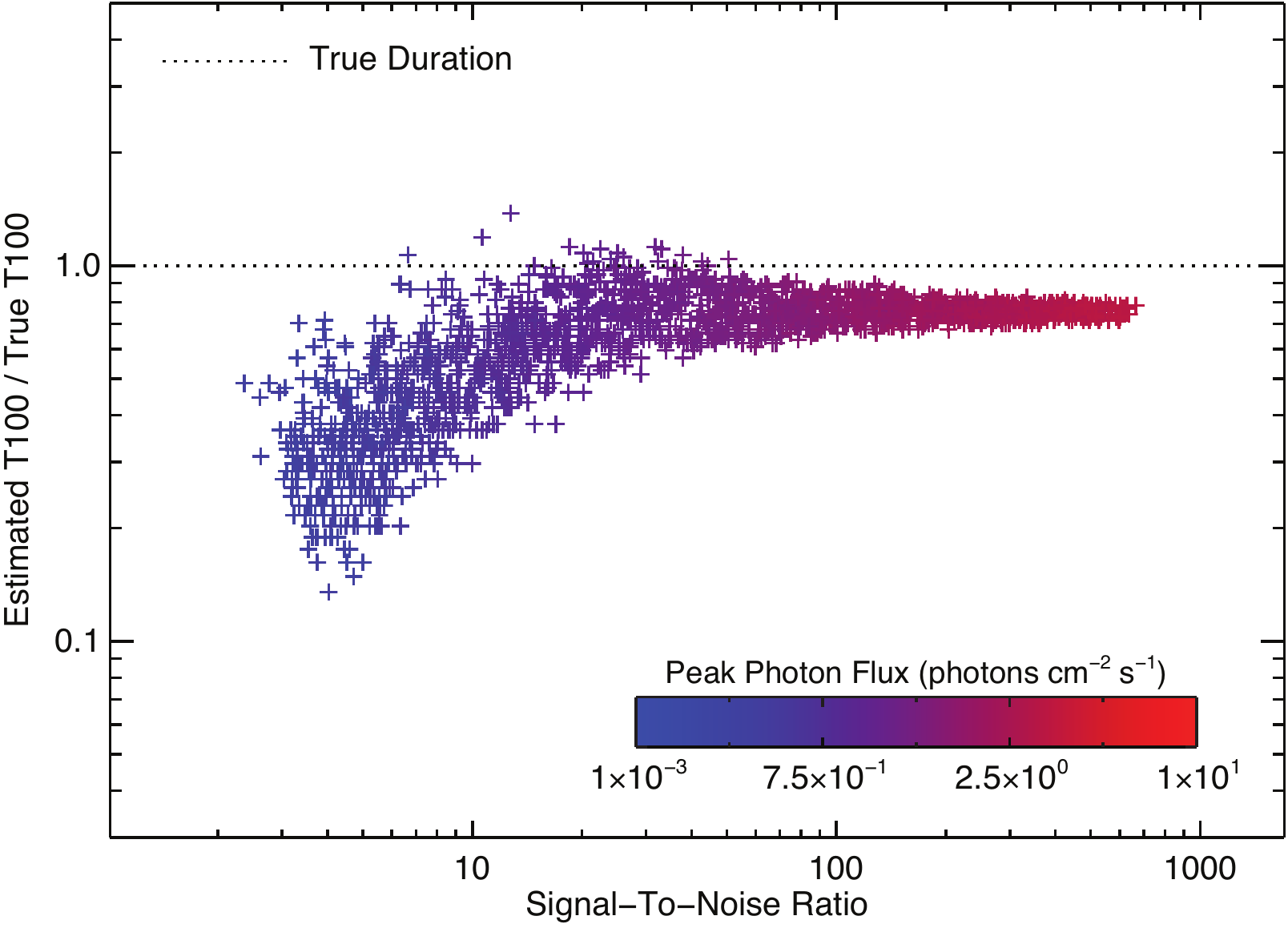}
\caption{The resulting fast rise exponential decay (FRED) pulse shape generated by the model by integrating the bolometric GRB spectrum at each time bin.}
\label{T100VsSNVsFlux}
\end{figure}

The systematic error introduced into our estimate of the observer frame duration has severe consequences for our subsequent estimates of the source frame duration and the isotropic equivalent energy $E_{\rm iso}$ of a GRB pulse.  By under-estimating the observer frame duration, any transformation into the source frame by dividing by $1+z$ only acts to further diverge our estimates of the source frame duration from their true values.  For the simulated pulses shown in Figure \ref{T100VsSNVsFlux}, this fundamental duration bias is largest for bursts detected near the instrument's detection threshold and can be as much as $90\%$ before the pulse is no longer detected by the instrument.  Such a large divergence between our estimated source frame duration and the true intrinsic duration result in large systematic error in our estimates of the burst's total energetics.

Our simulations show that, baring the existence of an intrinsic evolution in the duration of GRBs and redshift, we do not expect high redshift, single pulsed, GRBs to be characterized by extremely long pulse durations.  This conclusion agrees with the fact that evidence does not exist for the systematic broadening of GRB durations as a function of redshift in either the \emph{Swift} or \emph{Fermi} detected GRBs with known redshift.  The most distance GRBs to date, GRB~980425 at z = 9.4, GRB~090423 at $z \sim 8.1$ and GRB~080913A at $z \sim 6.7$, have measured durations of 5.5 sec \citep{Cucchiara11}, 10.3 sec \citep{Salvaterra09} and 8.1 sec \citep{Greiner09} respectively, giving them rest frame durations of $\sim0.52$ sec, $\sim1.13$ sec, and $\sim1.04$ sec.  Our simulations show that these values are actually lower limits to the true duration of these events and as such, so to are their estimated $E_{\rm iso}$ values.  

Although the results presented in Figure \ref{T100VsRedshift-ObserverFrame} show an increase in the observed duration of the five simulated pulses at low redshifts, a broad luminosity function, such as the one reported by \citet{Butler10}, will act to mask such time dilation signatures through the existence of intrinsically weak bursts at all redshifts.  Our simulations show that such a broad luminosity function effectively rules out the possibility that early studies successfully detected signatures of time dilation when considering the average duration of BATSE detected GRBs as a function of peak flux \citep{Norris94}.  This is essentially the same argument presented by \citet{Band94}, years before the first measurement of redshifts associated with GRB afterglows and their host galaxies.

The fundamental duration bias presented here has broad consequences of our understand of the energetics of GRBs, as the true $E_{\rm iso}$ distribution may extend to higher energies than had previously been estimated \citep{Butler10,Kocevski08}.  It also implies that attempts to predict the time of jet $t_{\rm jet}$ breaks in afterglow light curve, usually performed by measuring $E_{\rm iso}$ and assuming a canonical collimation corrected energy $E_{\gamma}$, may also be suspect.  If $E_{\rm iso}$ values for low signal-to-noise ratio bursts are truly lower limits to the true energy, then their associated  $t_{\rm jet}$ values are likewise lower limits and their jet break may actually occur at a point when the afterglow has faded beyond detectability.

Although measurements of $E_{\rm iso}$ may suffer from the systematic bias in duration measurements, estimate of a burst's peak isotropic equivalent luminosity $L_{\rm iso,pk}$ should not be effected, as long as a measurement of  $E_{\rm pk,obs}$ allows for a proper k-correction of the burst's underlying spectra.  This would mean that although $E_{\rm iso}$ evolution studies may in fact be biased at high redshift where a greater percentage of detected bursts have low signal-to-noise ratios, no such bias should exist when considering $L_{\rm iso}$.  Therefore, studies finding the lack of luminosity evolution \citep{Butler10} as a function of redshift would not be effected, although the similar attempts to quantify the evolution of $E_{\rm iso}$ may be suspect.

Ultimately, it is the shape of GRB pulses that leads to the fundamental duration bias observed in this analysis.  In general, signals with high kurtosis will suffer a greater degree of bias in their duration estimate, whereas a signals with low kurtosis will suffer no bias at all.  For example, a duration bias would not manifest for a signal exhibiting a step function profile.  The duration of the step function would simply increase as a function of $1+z$ until it fell below the instrument's detection threshold. 

Although our simulations were performed using a BATSE data response matrix, these results should generally apply to any flux limited instrument that collects time series data.  The results presented in Figure \ref{T100VsSNVsFlux} indicate that a burst's observed signal-to-noise ratio can provide a useful proxy to determine the severity of the bias in the estimates of burst duration.  In the simulations presented in Figure \ref{T100VsSNVsFlux}, any bursts detected by BATSE with a signal to noise less than $\sim 25$ should be considered heavily biased.  In a similar fashion, simulations of a large number of GRBs of varying luminosity, duration, and source frame spectra for different spacecraft may aid in determining the instrument specific signal-to-noise ratio below which duration estimates may be considered suspect.
 
The existence of the classic bimodal duration distribution reported in \citet{Kouveliotou93} would seem to limit the degree to which this duration bias can produce arbitrarily short durations for intrinsically long duration GRBs, despite that we have shown that GRB durations can be underestimated in some cases as much as 80$-$90$\%$.  This apparent contradiction can be reconciled by considering that most GRBs are in fact not single pulsed events, but comprise multiple emission periods.  Contrary to the effects that increasing redshift have on the observed duration of individual pulses, the duration between pulses, i$.$e$.$ the quiescent periods, should in fact increase.  Therefore, although long GRBs that exhibit a single, low signal-to-noise, pulse may result in observed durations that would allow them to be confused with the short GRB population, a majority of long GRBs would not suffer this fate even though their durations would still be under-estimated.  None-the-less, our work demonstrates that attempts at distinguishing between long and short GRBs based on a burst's temporal properties alone \citep{Fynbo06c, Ofek07}) cannot be trusted and additional information about the burst (e.g. spectral lag, host galaxy properties, hardness ratios) must be brought to bare.
 
A profile exhibiting narrow pulses separated by significant quiescent periods, may in fact be the tell tail sign of a high redshift GRB.  Although there are undoubtably low redshift GRBs which intrinsically exhibit such temporal profiles, additional spectral information could be used in order to distinguish these bursts from their high redshift counterparts,  Therefore, a burst's hardness as measured by an instrument with broad-band energy coverage, such as the Fermi-GBM, combined with its observed temporal profile and signal-to-noise ratio could play an important role in distinguishing high redshift events.  

Finally, we conclude by predicting that the average peak-to-peak time for a large number of multi-pulsed GRBs as a function of redshift may eventually provide the evidence for time dilation that has so far eluded detection.



\section{Conclusions} \label{sec:Conclusions}

By simulating the effects of time dilation on the evolution of the prompt gamma-ray spectra, we have found that:

\indent $\bullet$ The observer frame duration of individual GRB pulses does not always increase as a function of redshift as one would expect from the cosmological expansion of a Friedman-Lema\^{\i}tre-Robertson-Walker Universe. \\
\indent $\bullet$ The observed duration of individual pulses can actually decrease with increasing redshift as only the brightest portion of a high redshift GRB's light curve become accessible to the detector.  \\
\indent $\bullet$ The net result of this fundamental duration bias is that all estimates of duration and energetics for high redshift GRBs can only be taken as lower limits to their true values.\\
\indent $\bullet$ Bursts detected near the instrument's detection threshold are effected the most by this bias and an instrument specific signal-to-noise ratio below which duration estimates may be considered suspect may be isolated through simulations. \\
\indent $\bullet$ Any attempts at distinguishing between long and short GRBs based on a burst's temporal properties alone should not be trusted.  Additional information about the burst its spectral lag, host galaxy properties, and hardness ratios must be used in such classification attempts. \\

We conclude that GRB temporal profiles that exhibit narrow pulses separated by significant quiescent periods, may in fact the tell tail sign of high redshift GRBs and predict that the average peak-to-peak duration for a large number of multi-pulsed GRBs as a function of redshift may eventually provide the evidence for time dilation that has so far eluded detection.

\acknowledgements

D.K. acknowledges financial support through the Fermi Guest Investigator program. This work was supported in part by the NASA Fermi Guest Investigator program under NASA grant number NNX10AD13G and the U.S. Department of Energy contract to the SLAC National Accelerator Laboratory under no. DE-AC3-76SF00515.

\bibliography{ms}

\begin{thebibliography}{18}
\expandafter\ifx\csname natexlab\endcsname\relax\def\natexlab#1{#1}\fi

\bibitem[{{Band} {et~al.}(1993){Band}, {Matteson}, {Ford}, {Schaefer},
  {Palmer}, {Teegarden}, {Cline}, {Briggs}, {Paciesas}, {Pendleton}, {Fishman},
  {Kouveliotou}, {Meegan}, {Wilson}, \& {Lestrade}}]{Band93}
{Band}, D., {Matteson}, J., {Ford}, L., {Schaefer}, B., {Palmer}, D.,
  {Teegarden}, B., {Cline}, T., {Briggs}, M., {Paciesas}, W., {Pendleton}, G.,
  {Fishman}, G., {Kouveliotou}, C., {Meegan}, C., {Wilson}, R., \& {Lestrade},
  P. 1993, \apj, 413, 281

\bibitem[{{Band}(1994)}]{Band94}
{Band}, D.~L. 1994, \apjl, 432, L23

\bibitem[{{Butler} {et~al.}(2010){Butler}, {Bloom}, \& {Poznanski}}]{Butler10}
{Butler}, N.~R., {Bloom}, J.~S., \& {Poznanski}, D. 2010, \apj, 711, 495

\bibitem[{{Campana} {et~al.}(2007){Campana}, {Tagliaferri}, {Malesani},
  {Stella}, {D'Avanzo}, {Chincarini}, \& {Covino}}]{Campana07}
{Campana}, S., {Tagliaferri}, G., {Malesani}, D., {Stella}, L., {D'Avanzo}, P.,
  {Chincarini}, G., \& {Covino}, S. 2007, \aap, 464, L25

\bibitem[{{Cucchiara} {et~al.}(2011){Cucchiara}, {Levan}, {Fox}, {Tanvir},
  {Ukwatta}, {Berger}, {Kr{\"u}hler}, {K{\"u}pc{\"u} Yolda{\c s}}, {Wu},
  {Toma}, {Greiner}, {Olivares}, {Rowlinson}, {Amati}, {Sakamoto}, {Roth},
  {Stephens}, {Fritz}, {Fynbo}, {Hjorth}, {Malesani}, {Jakobsson}, {Wiersema},
  {O'Brien}, {Soderberg}, {Foley}, {Fruchter}, {Rhoads}, {Rutledge}, {Schmidt},
  {Dopita}, {Podsiadlowski}, {Willingale}, {Wolf}, {Kulkarni}, \&
  {D'Avanzo}}]{Cucchiara11}
{Cucchiara}, A., {Levan}, A.~J., {Fox}, D.~B., {Tanvir}, N.~R., {Ukwatta},
  T.~N., {Berger}, E., {Kr{\"u}hler}, T., {K{\"u}pc{\"u} Yolda{\c s}}, A.,
  {Wu}, X.~F., {Toma}, K., {Greiner}, J., {Olivares}, F.~E., {Rowlinson}, A.,
  {Amati}, L., {Sakamoto}, T., {Roth}, K., {Stephens}, A., {Fritz}, A.,
  {Fynbo}, J.~P.~U., {Hjorth}, J., {Malesani}, D., {Jakobsson}, P., {Wiersema},
  K., {O'Brien}, P.~T., {Soderberg}, A.~M., {Foley}, R.~J., {Fruchter}, A.~S.,
  {Rhoads}, J., {Rutledge}, R.~E., {Schmidt}, B.~P., {Dopita}, M.~A.,
  {Podsiadlowski}, P., {Willingale}, R., {Wolf}, C., {Kulkarni}, S.~R., \&
  {D'Avanzo}, P. 2011, \apj, 736, 7

\bibitem[{{Fynbo} {et~al.}(2006){Fynbo}, {Watson}, {Th{\"o}ne}, {Sollerman},
  {Bloom}, {Davis}, {Hjorth}, {Jakobsson}, {J{\o}rgensen}, {Graham},
  {Fruchter}, {Bersier}, {Kewley}, {Cassan}, {Castro Cer{\'o}n}, {Foley},
  {Gorosabel}, {Hinse}, {Horne}, {Jensen}, {Klose}, {Kocevski}, {Marquette},
  {Perley}, {Ramirez-Ruiz}, {Stritzinger}, {Vreeswijk}, {Wijers}, {Woller},
  {Xu}, \& {Zub}}]{Fynbo06c}
{Fynbo}, J.~P.~U., {Watson}, D., {Th{\"o}ne}, C.~C., {Sollerman}, J., {Bloom},
  J.~S., {Davis}, T.~M., {Hjorth}, J., {Jakobsson}, P., {J{\o}rgensen}, U.~G.,
  {Graham}, J.~F., {Fruchter}, A.~S., {Bersier}, D., {Kewley}, L., {Cassan},
  A., {Castro Cer{\'o}n}, J.~M., {Foley}, S., {Gorosabel}, J., {Hinse}, T.~C.,
  {Horne}, K.~D., {Jensen}, B.~L., {Klose}, S., {Kocevski}, D., {Marquette},
  J.-B., {Perley}, D., {Ramirez-Ruiz}, E., {Stritzinger}, M.~D., {Vreeswijk},
  P.~M., {Wijers}, R.~A.~M., {Woller}, K.~G., {Xu}, D., \& {Zub}, M. 2006,
  \nat, 444, 1047

\bibitem[{{Galama} {et~al.}(1999){Galama}, {Vreeswijk}, {van Paradijs},
  {Kouveliotou}, {Augusteijn}, {Patat}, {Heise}, {in 't Zand}, {Groot},
  {Wijers}, {Pian}, {Palazzi}, {Frontera}, \& {Masetti}}]{Galama99}
{Galama}, T.~J., {Vreeswijk}, P.~M., {van Paradijs}, J., {Kouveliotou}, C.,
  {Augusteijn}, T., {Patat}, F., {Heise}, J., {in 't Zand}, J., {Groot}, P.~J.,
  {Wijers}, R.~A.~M.~J., {Pian}, E., {Palazzi}, E., {Frontera}, F., \&
  {Masetti}, N. 1999, \aaps, 138, 465

\bibitem[{{Gehrels} {et~al.}(2004){Gehrels}, {Chincarini}, {Giommi}, {Mason},
  {Nousek}, {Wells}, {White}, {Barthelmy}, {Burrows}, {Cominsky}, {Hurley},
  {Marshall}, {M{\'e}sz{\'a}ros}, {Roming}, {Angelini}, {Barbier}, {Belloni},
  {Campana}, {Caraveo}, {Chester}, {Citterio}, {Cline}, {Cropper}, {Cummings},
  {Dean}, {Feigelson}, {Fenimore}, {Frail}, {Fruchter}, {Garmire}, {Gendreau},
  {Ghisellini}, {Greiner}, {Hill}, {Hunsberger}, {Krimm}, {Kulkarni}, {Kumar},
  {Lebrun}, {Lloyd-Ronning}, {Markwardt}, {Mattson}, {Mushotzky}, {Norris},
  {Osborne}, {Paczynski}, {Palmer}, {Park}, {Parsons}, {Paul}, {Rees},
  {Reynolds}, {Rhoads}, {Sasseen}, {Schaefer}, {Short}, {Smale}, {Smith},
  {Stella}, {Tagliaferri}, {Takahashi}, {Tashiro}, {Townsley}, {Tueller},
  {Turner}, {Vietri}, {Voges}, {Ward}, {Willingale}, {Zerbi}, \&
  {Zhang}}]{Gehrels04}
{Gehrels}, N., {Chincarini}, G., {Giommi}, P., {Mason}, K.~O., {Nousek}, J.~A.,
  {Wells}, A.~A., {White}, N.~E., {Barthelmy}, S.~D., {Burrows}, D.~N.,
  {Cominsky}, L.~R., {Hurley}, K.~C., {Marshall}, F.~E., {M{\'e}sz{\'a}ros},
  P., {Roming}, P.~W.~A., {Angelini}, L., {Barbier}, L.~M., {Belloni}, T.,
  {Campana}, S., {Caraveo}, P.~A., {Chester}, M.~M., {Citterio}, O., {Cline},
  T.~L., {Cropper}, M.~S., {Cummings}, J.~R., {Dean}, A.~J., {Feigelson},
  E.~D., {Fenimore}, E.~E., {Frail}, D.~A., {Fruchter}, A.~S., {Garmire},
  G.~P., {Gendreau}, K., {Ghisellini}, G., {Greiner}, J., {Hill}, J.~E.,
  {Hunsberger}, S.~D., {Krimm}, H.~A., {Kulkarni}, S.~R., {Kumar}, P.,
  {Lebrun}, F., {Lloyd-Ronning}, N.~M., {Markwardt}, C.~B., {Mattson}, B.~J.,
  {Mushotzky}, R.~F., {Norris}, J.~P., {Osborne}, J., {Paczynski}, B.,
  {Palmer}, D.~M., {Park}, H.-S., {Parsons}, A.~M., {Paul}, J., {Rees}, M.~J.,
  {Reynolds}, C.~S., {Rhoads}, J.~E., {Sasseen}, T.~P., {Schaefer}, B.~E.,
  {Short}, A.~T., {Smale}, A.~P., {Smith}, I.~A., {Stella}, L., {Tagliaferri},
  G., {Takahashi}, T., {Tashiro}, M., {Townsley}, L.~K., {Tueller}, J.,
  {Turner}, M.~J.~L., {Vietri}, M., {Voges}, W., {Ward}, M.~J., {Willingale},
  R., {Zerbi}, F.~M., \& {Zhang}, W.~W. 2004, \apj, 611, 1005

\bibitem[{{Golenetskii} {et~al.}(1983){Golenetskii}, {Mazets}, {Aptekar}, \&
  {Ilinskii}}]{Golenetskii83}
{Golenetskii}, S.~V., {Mazets}, E.~P., {Aptekar}, R.~L., \& {Ilinskii}, V.~N.
  1983, \nat, 306, 451

\bibitem[{{Greiner} {et~al.}(2009){Greiner}, {Kr{\"u}hler}, {Fynbo}, {Rossi},
  {Schwarz}, {Klose}, {Savaglio}, {Tanvir}, {McBreen}, {Totani}, {Zhang}, {Wu},
  {Watson}, {Barthelmy}, {Beardmore}, {Ferrero}, {Gehrels}, {Kann}, {Kawai},
  {Yolda{\c s}}, {M{\'e}sz{\'a}ros}, {Milvang-Jensen}, {Oates}, {Pierini},
  {Schady}, {Toma}, {Vreeswijk}, {Yolda{\c s}}, {Zhang}, {Afonso}, {Aoki},
  {Burrows}, {Clemens}, {Filgas}, {Haiman}, {Hartmann}, {Hasinger}, {Hjorth},
  {Jehin}, {Levan}, {Liang}, {Malesani}, {Pyo}, {Schulze}, {Szokoly}, {Terada},
  \& {Wiersema}}]{Greiner09}
{Greiner}, J., {Kr{\"u}hler}, T., {Fynbo}, J.~P.~U., {Rossi}, A., {Schwarz},
  R., {Klose}, S., {Savaglio}, S., {Tanvir}, N.~R., {McBreen}, S., {Totani},
  T., {Zhang}, B.~B., {Wu}, X.~F., {Watson}, D., {Barthelmy}, S.~D.,
  {Beardmore}, A.~P., {Ferrero}, P., {Gehrels}, N., {Kann}, D.~A., {Kawai}, N.,
  {Yolda{\c s}}, A.~K., {M{\'e}sz{\'a}ros}, P., {Milvang-Jensen}, B., {Oates},
  S.~R., {Pierini}, D., {Schady}, P., {Toma}, K., {Vreeswijk}, P.~M., {Yolda{\c
  s}}, A., {Zhang}, B., {Afonso}, P., {Aoki}, K., {Burrows}, D.~N., {Clemens},
  C., {Filgas}, R., {Haiman}, Z., {Hartmann}, D.~H., {Hasinger}, G., {Hjorth},
  J., {Jehin}, E., {Levan}, A.~J., {Liang}, E.~W., {Malesani}, D., {Pyo},
  T.-S., {Schulze}, S., {Szokoly}, G., {Terada}, K., \& {Wiersema}, K. 2009,
  \apj, 693, 1610

\bibitem[{{Kocevski} \& {Butler}(2008)}]{Kocevski08}
{Kocevski}, D., \& {Butler}, N. 2008, \apj, 680, 531

\bibitem[{{Kocevski} {et~al.}(2003){Kocevski}, {Ryde}, \& {Liang}}]{Kocevski03}
{Kocevski}, D., {Ryde}, F., \& {Liang}, E. 2003, \apj, 596, 389

\bibitem[{{Kouveliotou} {et~al.}(1993){Kouveliotou}, {Meegan}, {Fishman},
  {Bhat}, {Briggs}, {Koshut}, {Paciesas}, \& {Pendleton}}]{Kouveliotou93}
{Kouveliotou}, C., {Meegan}, C.~A., {Fishman}, G.~J., {Bhat}, N.~P., {Briggs},
  M.~S., {Koshut}, T.~M., {Paciesas}, W.~S., \& {Pendleton}, G.~N. 1993, \apjl,
  413, L101

\bibitem[{{Liang} \& {Kargatis}(1996)}]{Liang96}
{Liang}, E., \& {Kargatis}, V. 1996, \nat, 381, 49

\bibitem[{{Meegan} {et~al.}(1992){Meegan}, {Fishman}, {Wilson}, {Horack},
  {Brock}, {Paciesas}, {Pendleton}, \& {Kouveliotou}}]{Meegan92}
{Meegan}, C.~A., {Fishman}, G.~J., {Wilson}, R.~B., {Horack}, J.~M., {Brock},
  M.~N., {Paciesas}, W.~S., {Pendleton}, G.~N., \& {Kouveliotou}, C. 1992,
  \nat, 355, 143

\bibitem[{{Norris} {et~al.}(1994){Norris}, {Nemiroff}, {Scargle},
  {Kouveliotou}, {Fishman}, {Meegan}, {Paciesas}, \& {Bonnel}}]{Norris94}
{Norris}, J.~P., {Nemiroff}, R.~J., {Scargle}, J.~D., {Kouveliotou}, C.,
  {Fishman}, G.~J., {Meegan}, C.~A., {Paciesas}, W.~S., \& {Bonnel}, J.~T.
  1994, \apj, 424, 540

\bibitem[{{Ofek} {et~al.}(2007){Ofek}, {Cenko}, {Gal-Yam}, {Fox}, {Nakar},
  {Rau}, {Frail}, {Kulkarni}, {Price}, {Schmidt}, {Soderberg}, {Peterson},
  {Berger}, {Sharon}, {Shemmer}, {Penprase}, {Chevalier}, {Brown}, {Burrows},
  {Gehrels}, {Harrison}, {Holland}, {Mangano}, {McCarthy}, {Moon}, {Nousek},
  {Persson}, {Piran}, \& {Sari}}]{Ofek07}
{Ofek}, E.~O., {Cenko}, S.~B., {Gal-Yam}, A., {Fox}, D.~B., {Nakar}, E., {Rau},
  A., {Frail}, D.~A., {Kulkarni}, S.~R., {Price}, P.~A., {Schmidt}, B.~P.,
  {Soderberg}, A.~M., {Peterson}, B., {Berger}, E., {Sharon}, K., {Shemmer},
  O., {Penprase}, B.~E., {Chevalier}, R.~A., {Brown}, P.~J., {Burrows}, D.~N.,
  {Gehrels}, N., {Harrison}, F., {Holland}, S.~T., {Mangano}, V., {McCarthy},
  P.~J., {Moon}, D.-S., {Nousek}, J.~A., {Persson}, S.~E., {Piran}, T., \&
  {Sari}, R. 2007, \apj, 662, 1129

\bibitem[{{Salvaterra} {et~al.}(2009){Salvaterra}, {Della Valle}, {Campana},
  {Chincarini}, {Covino}, {D'Avanzo}, {Fern{\'a}ndez-Soto}, {Guidorzi},
  {Mannucci}, {Margutti}, {Th{\"o}ne}, {Antonelli}, {Barthelmy}, {de Pasquale},
  {D'Elia}, {Fiore}, {Fugazza}, {Hunt}, {Maiorano}, {Marinoni}, {Marshall},
  {Molinari}, {Nousek}, {Pian}, {Racusin}, {Stella}, {Amati}, {Andreuzzi},
  {Cusumano}, {Fenimore}, {Ferrero}, {Giommi}, {Guetta}, {Holland}, {Hurley},
  {Israel}, {Mao}, {Markwardt}, {Masetti}, {Pagani}, {Palazzi}, {Palmer},
  {Piranomonte}, {Tagliaferri}, \& {Testa}}]{Salvaterra09}
{Salvaterra}, R., {Della Valle}, M., {Campana}, S., {Chincarini}, G., {Covino},
  S., {D'Avanzo}, P., {Fern{\'a}ndez-Soto}, A., {Guidorzi}, C., {Mannucci}, F.,
  {Margutti}, R., {Th{\"o}ne}, C.~C., {Antonelli}, L.~A., {Barthelmy}, S.~D.,
  {de Pasquale}, M., {D'Elia}, V., {Fiore}, F., {Fugazza}, D., {Hunt}, L.~K.,
  {Maiorano}, E., {Marinoni}, S., {Marshall}, F.~E., {Molinari}, E., {Nousek},
  J., {Pian}, E., {Racusin}, J.~L., {Stella}, L., {Amati}, L., {Andreuzzi}, G.,
  {Cusumano}, G., {Fenimore}, E.~E., {Ferrero}, P., {Giommi}, P., {Guetta}, D.,
  {Holland}, S.~T., {Hurley}, K., {Israel}, G.~L., {Mao}, J., {Markwardt},
  C.~B., {Masetti}, N., {Pagani}, C., {Palazzi}, E., {Palmer}, D.~M.,
  {Piranomonte}, S., {Tagliaferri}, G., \& {Testa}, V. 2009, \nat, 461, 1258

\end{thebibliography}

\end{document}